\documentclass[dvips]{article}

\usepackage{nolta,amsmath,amssymb,epsfig}

\newcommand{\diff}[2][{}]{\dfrac{\partial#1}{\partial #2}}
\newcommand{\rate}[3][{}]{\ensuremath{W^{#1}_{{#2}\rightarrow{#3}}}}
\newcommand{\konf}[1]{\ensuremath{\mathbf{#1}}}
\newcommand{\nob}[1][0]{\ensuremath{\mathbf{n}^{(#1)}}}
\newcommand{\Pbed}[2]{\ensuremath{\mathcal{P}\bigl(#1\,\big|\,#2\bigr)}}
\newcommand{\Rate}[2][{}]{\ensuremath{{W}^{#1}_{#2}}}
\newcommand{\proc}[1]{\texttt{#1}}
\newcommand{\intoa}[2]{\ensuremath{(#1,#2\,]}}
\newcommand{\mec}[1]{\ensuremath{\mathbf{#1}}}    
\newcommand{\Ovx}[1]{\ensuremath{\mathrm{O}(#1)}}
\newcommand{\mkomma}{\;,\quad}  
\newcommand{\mpunkt}{\;\;.}
\newcommand{\intao}[2]{\ensuremath{[\,#1,#2)}}

\begin{document}

\title{A Fast Algorithm for High-Dimensional Markov Processes
with Finite Sets of Transition Rates}

\author{Hans Ekkehard Plesser\textsuperscript{1}\ 
	and Dietmar Wendt\textsuperscript{2}\\
\\
\textsuperscript{1}\parbox[t]{12cm}{Laboratory for Neural Modeling, FRP, RIKEN,
2-1~Hirosawa,\\ Wako-shi, Saitama~351-01,~Japan
(plesser@yugiri.riken.go.jp)}\\
\textsuperscript{2}\parbox[t]{12cm}{Institut f\"ur Theoretische
Physik, RWTH Aachen,
52056~Aachen,\\Germany (D.Wendt@physik.rwth-aachen.de)}
}

\maketitle

\raisebox{7cm}[0cm][0cm]{\parbox[t][0cm][b]{\textwidth}{\textsf{\small%
   \hspace*{\fill}1996 International Symposium on Nonlinear Theory and its
      Applications\\
   \hspace*{\fill}NOLTA '96, Katsurahama-so, Kochi, Japan, Oct.~7--9, 1996, pp.~249--252}}}

\subsection*{\centering \sc Abstract}
The discrete class algorithm presented in this paper is an efficient
simulation tool for stochastic processes governed by a reasonably
small set of transition rates.  The algorithm is presented, its
performance compared to prevailing methods and applications to
epitaxial growth and neuronal models are sketched.

\section{Introduction}
Stochastic processes play a crucial role in many fields of science and
technology and have received much attention ever since the
ground-breaking work by Einstein, Smoluchowski and others at the
beginning of the
century~\cite{Einstein:Unte()(1922),Smoluch:Abha()(1923)}.  While many
low-dimensional stochastic processes can be treated
analytically~\cite{Gardine:Hand()(1985),Kampen:Stoc()(1992)}, this is
no longer the case for spatially extended, high-dimensional systems,
such as diffusion-limited reaction-diffusion systems, epitaxial
growth, population dynamics or neuronal
interactions~\cite{Wendt:1995(541),Ratsch:1993(3194),Engbert:1994(1147),%
Ohira:1993(2259)}.

The unifying feature of all these systems is that their development
in time is given by a master equation 
\begin{multline*}
  \diff{t}\Pbed{\konf{n},t}{\nob} =\\
  \sum_{\konf{n'}}
      \rate{\konf{n'}}{\konf{n}} \Pbed{\konf{n'}, t}{\nob}
    	-\rate{\konf{n}}{\konf{n'}}  \Pbed{\konf{n},t}{\nob}
\end{multline*}
for the probability of the system to be in state $\konf{n}$ at time $t$
if it was in state $\nob$ at time $t^{(0)}$.  Here,
$\rate{\konf{n}}{\konf{n'}}$ are transition rates and $\konf{n}$ is a
vector in an $m$-dimensional discrete state space.  Since the
master equation can rarely be solved analytically, efficient
simulation methods for the generation of trajectories obeying the
equation are of tantamount importance.

In the next section, we present the highly efficient \emph{discrete
class algorithm} (DCA) for the simulation of systems governed by a
reasonably small set of different transistion rates.  
In section~3 we apply the
algorithm to a simple model of epitaxial growth and demonstrate the
speed-up compared to prevailing methods.  Finally, in
section~4, we show how the DCA can be used to study large
neural networks.

\section{The Discrete Class Algorithm}\label{sec:dca}
The discrete class algorithm is an extension of the minimal process
method~\cite{Karlin:Firs()(1975)} introduced by
Gillespie~\cite{Gillesp:1976(403)}.  This elegant
algorithm proceeds from state $\konf{n}$ at $t$ to $\konf{n'}$ at
$t'=t+\tau$ as follows:
\begin{enumerate}
\item Calculate the total rate for leaving state $\konf{n}$:\\
  \mbox{$\Rate{\konf{n}} = \sum_{\konf{n'}}\rate{\konf{n}}{\konf{n'}}$}.
\item Determine the (exponentially distributed) time step
  \mbox{$\tau = - \ln\proc{rnd}\intoa{0}{1} / \Rate{\konf{n}}$}.
\item Choose a new state \konf{n'} with probability \\
  \mbox{$p_{\mec{n}}(\konf{n'})=\rate{\konf{n}}{\konf{n'}}/\Rate{\konf{n}}$}. 
\end{enumerate}

The efficiency of this algorithm hinges on the efficient
implementation of step~3, i.e.\ the selection of the new state
\konf{n'} from the probabililty distribution $p_{\mec{n}}(\konf{n'})$
depending on the current state \konf{n}.  For an $m$-dimensional
problem, the number of states \konf{n'} accessible from \konf{n}
($\rate{\konf{n}}{\konf{n'}} > 0$) will be $M \sim \Ovx{m}$.
To see this, consider a reaction-diffusion system with two
particle species $\mathrm{A}$ and $\mathrm{B}$ and one chemical
reaction $\mathrm{A}+\mathrm{B}\rightarrow\oslash$, modelled on a grid
of $L\times L$ cells.  Then, the state space is $m=2L^2$-dimensional and
as long as all cells contain at least one $\mathrm{A}$ and one
$\mathrm{B}$ particle each, at every time step one out of
$M=(4+4+1)\cdot L^2$ possible events has to be chosen:  From any one
of the $L^2$ cells, either an $A$
or a $B$ particle diffuses to any one of the four nearest neighbors or
a reaction occurs in it.
  Therefore, linear
selection schemes requiring an effort of \Ovx{M} additions per time
step are utterly unsuitable for large systems, as are rejection methods,
which are efficient only if $p_{\mec{n}}(\konf{n'})$ is restricted to
a small interval~\cite{Fricke:1995(277)}.  Methods employing
binary trees for step~3 still require an effort of
$\Ovx{\log_2 M}$ per time step~\cite{Blue:1995(R867)}.

A sophisticated method using a logarithmic classification scheme for
step~3 has been introduced a few years
ago~\cite{Fricke:1991(277),Matias:1993}.  This method
yields a computational effort for step 3 independent of $M$, i.e.\
independent of the size of the system, provided the transition rates
\rate{\konf{n}}{\konf{n'}} are independent of $M$ (for a more detailed
discussion see~\cite{Fricke:1995(277)}). The algorithm involves some overhead
though, and for systems spanning a very large range of transition
rates ($ > \Ovx{10^{14}}$), the algorithm slows down slightly due to
computational precautions required to avoid round-off errors.

The discrete class algorithm is similar in spirit to the logarithmic
classes, but specifically aimed at systems with a reasonably small,
discrete set of transition rates, i.e.
\begin{equation*}
   \rate{\konf{n}}{\konf{n'}} \in \{r_1, \dots, r_K\}\mkomma 
	K\lesssim 50 \mpunkt
\end{equation*}
While this may seem a to be a strong restriction at first, models of
epitaxial growth~\cite{Maksym:1988(594)} fulfill these restrictions
as well as some neuronal models.

\vspace{3mm}

The DCA implements step~3 of the minimal process
method as follows.  Each possible transition event
\mbox{$\konf{n}\rightarrow\konf{n'}$} is assigned to one of $K$ classes
according to its rate
\begin{equation*}
  D_{\nu} = \{\konf{n}\rightarrow\konf{n'}\,|\,\rate{\konf{n}}{\konf{n'}} 
          = r_{\nu} \} \mkomma \nu \in 1,\dots,K \quad .
\end{equation*}
Thus the rate of events in class $D_{\nu}$ is given by
\begin{equation*}
  R_{\nu} = \sum_{\konf{n'}\in D_{\nu}} \rate{\konf{n}}{\konf{n'}} 
	= \| D_{\nu} \| \, r_{\nu} \mkomma
\end{equation*}
where $\| D_{\nu} \|$ is the number of events in $D_{\nu}$, while
the total rate of events is given by 
\begin{equation*}
  \Rate{\konf{n}}= \sum_{\nu=1}^{K} R_{\nu} \mpunkt
\end{equation*}
Furthermore, within a class each event occurs with equal probability.
The selection step~3 is thus split into two substeps:
\begin{itemize}
\item[3a.] Choose a class $D_{\nu}$ with probability $R_{\nu} /
  \Rate{\konf{n}}$ by linear selection, i.e.\ for a \hbox{$\rho =
  \proc{rnd}\intao{0}{\Rate{\konf{n}}}$} select that class $\nu$ for
  which 
  \begin{equation*}
	 \sum_{i=1}^{\nu-1} R_i \leq \rho < \sum_{i=1}^{\nu} R_i\;.
  \end{equation*}
\item[3b.] Select the new state \konf{n'} from class
  $D_{\nu}$ at random.
\end{itemize}

The linear selection in step~3a requires drawing a single, uniformly
distributed random number and \Ovx{K} additions, independent of the
size of the system, while step~3b requires drawing another uniformly
distributed random number.  Thus, the efficiency of the selection
algorithm does not depend on the size of the system under study.  As
the number of classes is assumed to be small, the total rate
$\Rate{\konf{n}}$ can be calculated at every step, keeping numeric
inaccuracies to a minimum.
   \begin{figure}
      \begin{center}
         \setlength{\unitlength}{1mm}
         \begin{picture}(75,55)(0,0)
            \put( 2, 1){\epsfig{file=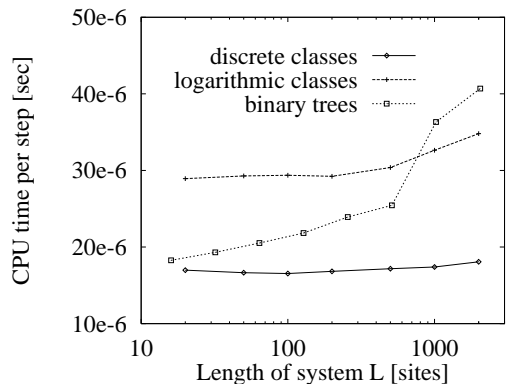,
                                width=72mm,clip=}}
         \end{picture}
         \parbox{0.9\linewidth}{\caption{CPU time required per
	time step for different simulation algorithms.  All
        simulations were performed on an SGI Indy workstation with
	192 megabyte of memory.
	}\label{fig:times}} 
      \end{center}
   \end{figure}

\section{Comparison with other methods}\label{sec:mbe}
To demonstrate the performance of the DCA compared to the logarithmic
classes~\cite{Wendt:1995(541),Fricke:1995(277)} and other
state-of-the art methods such as binary trees~\cite{Blue:1995(R867)},
let us consider a simple model of epitaxial growth based
on~\cite{Ratsch:1993(3194)}:
\begin{itemize}
  \item the substrate is an $L\times L$ lattice;
  \item each lattice site is occupied by one adatom or none;
  \item in an initial phase, $N < L\times L$ adatoms are deposited,
        which cannot evaporate;
  \item an adatom with all four next neighbor sites occupied cannot move;
  \item all other adatoms diffuse to next neighbor sites with rates
        \begin{equation*}\displaystyle
        w_n = \tfrac{2k_B T}{h} e^{-E_S/k_B T}\, e^{-nE_N/k_B T}\mpunkt
        \end{equation*}
\end{itemize}
Here, $n$ is the number of occupied next-neighbor sites, $h$, $k_B$
are Planck's and Boltzmann's constants, $T$ is temperature and $E_S$,
$E_N$ are material-dependent energies characterizing adatom-substrate
and adatom-adatom interactions, respectively; both are on the order of
$1\mathrm{eV}$.  Thus, after the deposition phase, the system is
determined by just five different rates which are at
$T=600\mathrm{K}$: $w_0 = 3.0\cdot10^2$, $w_1 = 1.2\cdot10^{-6}$, $w_2
= 4.8\cdot10^{-15}$, $w_3 = 1.9\cdot10^{-23}$ and $w_4 \equiv 0$.

Figure~\ref{fig:times} shows the CPU~time required per step for the
simulation of this model for different lattice sizes using the discrete
class, the logarithmic class and the binary tree algorithm.  This
demonstrates clearly the superiority of the discrete classes to the
other algorithms in terms of absolute times as well as the
size-independence of efficiency.  The minuscule increase in CPU~time
for the DCA at very large lattices is due to cache effects, i.e.\
shortcomings of the hardware; for a detailed discussion,
see~\cite{pre}.

   \begin{figure}
      \begin{center}
         \setlength{\unitlength}{1mm}
         \begin{picture}(75,75)(0,0)
	     \put(15,25){\texttt{\large see retina.gif}}
         \end{picture}
         \parbox{0.9\linewidth}{\caption{Propagation of waves on the
	   retina. Grey level indicates time since last firing, 
	   black being most recent.
	}\label{fig:retina}} 
      \end{center}
   \end{figure}
\section{A Neuronal Model}\label{sec:neuron}
A crucial problem in modeling the signal processing by neuronal
networks is the enormous number of neurons involved even in simple
tasks.  Typically, though, only a small number of neurons will respond
to any one stimulus presented e.g.\ to the eye or the ear.  
The DCA is well suited for the simulation of such largely ``dormant''
systems, since it automatically ``focuses'' on active regions of the
system under study.  

To demonstrate the applicability of the DCA to neuronal studies, we
have formulated the following model, which is essentially a simplified 
type of Stein's model neuron~\cite{Tuckwell:1989}.
\begin{itemize}
  \item At $t=0$, each neuron $j$ has the resting membrane 
        potential $v_j(0)=0$.
  \item The membrane potential $v_j$ is governed by the equation
        $\mathrm{d}v_j/\mathrm{d}t = f_s^{(j)}(t) + f_p^{(j)}(t)$.
  \item All input $f_s^{(j)}$, $f_p^{(j)}$ consists of delta-spikes, 
	i.e.~an input event at time $T$ corresponds to the transition 
	$v_j(T-) \rightarrow v_j(T+) = v_j(T-)+1$.
  \item $f_p^{(j)}(t)$ is Poissonian noise with rate $1/\tau_p$.
  \item $f_s^{(j)}(t)$ is synaptic input from other neurons.
  \item As the potential reaches a threshold, $v_j(t)=\Theta$, 
        neuron $j$ fires a spike after an average waiting time $\tau_f$,
        which is transmitted to all $k_j$ neurons receiving input from
        $j$; neuron $j$ is reset to an absolute refractory state.
  \item All input is ignored in the refractory state and the neuron
        returns to the resting state $v_j(t)=0$ with rate $1/\tau_r$.
\end{itemize}
This model is obviously very well suited for the DCA, since it is
governed by only three different rates: $1/\tau_p$, $1/\tau_f$ and
$1/\tau_r$.  In order to model spontaneous retinal waves as have been
observed in newborn ferrets~\cite{Feller:1996(1182)}, we have
simulated a grid of neurons with strongly localized synaptic
connections.  The network studied had $512\times512$ neurons and some
$6.7\cdot10^6$ synapses, the threshold was set to $\Theta=7$.  The
simulation was stopped after $1.5$~million spikes had been generated,
which required only 80 seconds of CPU time on an SGI~Indy workstation.
Figure~\ref{fig:retina} shows a typical state of
activity.

\section{Conclusions}
The DCA algorithm presented here is a powerful tool for the study of a 
large class of stochastic systems and should foster research in these
fields.  The extension of the epitaxial model towards more complex
phenomena is straightforward.  

The neuron model presented above is most likely too simplistic to
further our understanding of real neuronal systems, but we are
presently working on a faithful implementation of Stein's model.
Preliminary results indicate that leak currents and inhibitory inputs
can be included.  Inclusion of arbitrary synaptic weights, though,
might necessitate recourse to the more generally applicable
logarithmic class algorithm.

Note that the effective implementation of the DCA requires
sophisticated data structures, similar to those described
in~\cite{Fricke:1995(277)}.  Source code that can be integrated in
simulation software via an easy to use interface is available from the
authors upon request.

\section*{Acknowledgement}
Hans E. Plesser acknowledges partial financial support by
Studien\-stif\-tung des deutschen Volkes.

\sloppy

\end{document}